\documentclass[sigconf, nonacm]{acmart}

\AtBeginDocument{%
  }

\usepackage{enumitem}
\setlist[enumerate]{leftmargin=2.6em}
\usepackage{multirow}
\usepackage{pifont}
\usepackage[ruled,lined,linesnumbered]{algorithm2e}
\usepackage{listings}
\usepackage{float}
\newfloat{listing}{htbp}{lol}
\floatname{listing}{Code}
\newcommand{\cmark}{\ding{51}}
\newcommand{\xmark}{\ding{55}}
\newcommand{\pmark}{$\sim$}

\begin{document}

\title{SEAR: Schema-Based Evaluation and Routing for LLM Gateways}

\author{Zecheng Zhang}
\affiliation{%
  \institution{Strukto.AI}
  \country{USA}}
\email{zecheng@strukto.ai}

\author{Han Zheng}
\affiliation{%
  \institution{Infron.AI}
  \country{USA}}
\email{andrew.zheng@infron.ai}

\author{Yue Xu}
\affiliation{%
  \institution{Infron.AI}
  \country{USA}}
\email{lawrence.xu@infron.ai}

\begin{abstract}
Evaluating production LLM responses and routing requests across
providers in LLM gateways requires fine-grained quality signals and
operationally grounded decisions.
To address this gap, we present SEAR, a schema-based evaluation and
routing system for multi-model, multi-provider LLM gateways.
SEAR defines an extensible relational schema covering both LLM
evaluation signals (context, intent, response characteristics, issue
attribution, and quality scores) and gateway operational metrics
(latency, cost, throughput), with cross-table consistency links across
around one hundred typed, SQL-queryable columns.
To populate the evaluation signals reliably, SEAR proposes
self-contained signal instructions, in-schema reasoning, and
multi-stage generation that produces database-ready structured
outputs. Because signals are derived through LLM reasoning rather than
shallow classifiers, SEAR captures complex request semantics, enables
human-interpretable routing explanations, and unifies evaluation and
routing in a single query layer.
Across thousands of production sessions, SEAR achieves strong signal
accuracy on human-labeled data and supports practical routing decisions,
including large cost reductions with comparable quality.

\end{abstract}



\maketitle

\section{Introduction}

As production LLM traffic grows, evaluating agentic systems at scale
remains challenging.
Workloads are diverse, tasks evolve over time, and failures are often
concentrated in specific traffic subsets rather than uniformly
distributed.
In particular, production deployments span domains such as technology,
healthcare, finance, and education, with multi-turn conversations,
reference documents, and multi-modal inputs across varying complexity
levels.
As a result, no single model or provider is optimal for all cases, and
costs differ by orders of magnitude.

Teams therefore rely on multi-model compositions that route cheaper
models to simpler tasks and stronger models to harder ones
~\cite{dinghybrid,hu2024routerbench}.
Recent empirical studies show that price alone is not sufficient for
model selection~\cite{aubakirova2026stateofai}, underscoring the need
for fine-grained, quality-aware routing signals.
However, this setup introduces compounding complexity, from model
assignment and cross-provider quality evaluation to continual
reassessment as models evolve.
This complexity is increasing with agentic inference, where
reasoning-model adoption, tool-call usage, and context length are
rising sharply~\cite{aubakirova2026stateofai}.
Meanwhile, public benchmarks saturate quickly and often fail to reflect
real-world usage~\cite{yao2025secondhalf,fodor2025benchmarklimitations,chiang2024chatbotarena},
leaving teams to rely on manual spot checks, small internal benchmarks,
and proxy metrics.

For quality assessment, LLM-as-Judge evaluation, where one LLM
evaluates the output of another, has become the dominant paradigm for
production evaluation at scale~\cite{zhengjudging}.
Current approaches fall into three broad categories, namely
unstructured methods that produce free-text commentary with post-hoc
parsing for scores, single-score evaluators that collapse all quality
dimensions into one holistic rating, and rubric-based evaluators that apply manually designed,
fixed criteria across a limited number of
dimensions~\cite{kim2024prometheus}.
Commercially, template-based evaluator pipelines~\cite{langfuse2024}
let each team define custom scoring templates per use case.
Each approach has significant limitations.
Unstructured outputs are difficult to aggregate or query across
sessions at scale; single scores prevent drill-down into specific
failure modes; rubric-based evaluators produce a small fixed set of
scores that do not decompose into fine-grained, per-signal diagnostics;
and template-based pipelines fragment across teams, lack type
enforcement, and store results as untyped score-reasoning pairs that
are also difficult to aggregate.

On the routing side, existing approaches train routers with various
optimization objectives to select among
models~\cite{ong2024routellm,fenggraphrouter,zhang2025router,daipersonalizedrouter},
but their decisions remain black-box, and they provide a
recommendation without interpretable, signal-level explanations of why
a model suits a given task.
Interpretability is especially critical in production gateway settings,
where routing changes directly affect live services and teams need clear
justifications before modifying model assignments.
Moreover, production teams must balance not only performance but also
provider choice, cost, latency, and throughput, requiring routing logic
that exposes these trade-offs explicitly.
Deploying black-box routing decisions directly to live traffic is
therefore high-risk, and in practice many teams prefer asynchronous
policy updates derived from logged traffic, reviewed and validated
offline before deployment.

Our system, \textbf{S}chema-Based \textbf{E}valuation \textbf{a}nd \textbf{R}outing (SEAR), addresses
these gaps.

At its core, SEAR uses an LLM judge to generate interlinked relational
tables from each LLM request session, capturing around one hundred typed
signals across the full request lifecycle, from user intent and LLM
response characteristics to issue attribution and quality scores.
By deriving signals through LLM reasoning rather than shallow
classifiers, SEAR captures complex request semantics that heuristic
extractors miss, produces human-interpretable routing explanations
grounded in per-signal evidence, and unifies evaluation and routing
in a single queryable data layer.
Rather than requiring free-text parsing or custom per-team templates,
every signal is a typed, SQL-queryable column.
To produce these tables reliably, the judge follows a
schema-driven generation process that decomposes the task along
foreign-key dependencies and uses self-contained signal instructions and
in-schema reasoning to emit structured outputs in one call per table.
Alongside these evaluation signals, the gateway layer logs operational
metrics such as latency, throughput, cost, and error rates for every
request. Because both live in the same SQL-queryable data layer, teams
can jointly analyze response quality and operational performance through
standard queries. As signals accumulate, they refine routing decisions and routed traffic
produces new signals, forming a data flywheel supported by asynchronous
judging off the serving path.

In summary, our main contributions are:
\begin{enumerate}[leftmargin=2.6em]
  \item A data-driven system combining SQL-queryable LLM evaluation
  records with gateway operational metrics for flexible quality
  analysis, diagnosis, and routing.
  \item An extensible relational evaluation schema with cross-table
  consistency checks covering the full LLM request lifecycle.
  \item A schema-driven judge using self-contained signal
  instructions, in-schema reasoning, and multi-stage generation for
  reliable structured generation at this scale.
  \item Validation on production LLM gateway traffic demonstrating high
  signal accuracy and practical, cost-effective routing decisions.
\end{enumerate}

\section{Related Work}

\subsection{LLM-as-Judge Evaluation}
Using LLMs to evaluate LLM outputs has become a practical alternative
to human annotation at scale~\cite{zheng2023judging,zhu2023judgelm,wang2023chatgptevaluator}.
Methods range from single-score evaluators~\cite{liu2023geval,fu2023gptscore},
which are prone to scoring bias and self-inconsistency~\cite{li2025scoringbias,haldar2025ratingroulette},
to rubric-based approaches with multiple dimensions~\cite{zhong2022unieval,ye2023flask,kim2024prometheus},
though all rely on relatively few predefined dimensions or manual rubric design.

\subsection{LLM Structured Output}
Constraining LLM outputs to typed schemas is now supported natively by
major API providers~\cite{openai2024structuredoutputs,google2025geministructuredoutput,anthropic2025structuredoutputs}
and by constrained-decoding engines such as
Outlines~\cite{willard2023outlines} and XGrammar~\cite{dong2024xgrammar}.
However, format restrictions can degrade reasoning relative to free-form
generation~\cite{tam2024letmespeak,park2024gad,geng2025jsonschemabench}.

\subsection{Schema-Guided LLM Extraction}
Schema-aware extraction pipelines such as KARMA~\cite{karma2025} and
AOP~\cite{wang2025aop} coordinate multi-step extraction with
schema-aligned operators, but do not provide a shared signal space
across tasks for continuous quality analysis and routing.

\subsection{LLM Gateways and Monitoring}
LLM gateways provide unified serving and routing interfaces across model
providers~\cite{tensorzero2025}, while guardrails and observability
tools support policy enforcement, tracing, and cost
analytics~\cite{rebedea2023nemoguardrails,dong2024guardrailssurvey,langfuse2024}.
Evaluation-operations frameworks argue for continuous evaluation
throughout the agent lifecycle~\cite{dong2024agentops,xia2024eddops,zhao2025agentintheloop},
but in practice evaluation results are rarely connected to routing
policies without custom engineering.

\subsection{LLM Routing}
Cost-quality-aware routing selects a model per request under resource
constraints~\cite{varangot2025routingsurvey,hu2024routerbench}, with
approaches spanning preference-based~\cite{ong2024routellm,daipersonalizedrouter},
graph-based~\cite{fenggraphrouter}, cascade~\cite{chen2023frugalgpt,ding2024hybrid},
and RL-based~\cite{zhang2025router} methods.
Most routers optimize aggregate utility signals and do not expose
per-signal attributions, limiting transparency and signal-level
drift diagnosis in production gateway settings.
Signal-driven routers~\cite{liu2025vllmsemanticrouter} compose routing
policies from heuristic and classifier-extracted features, but are
limited to shallow signal extraction.

\section{The SEAR Framework}

\begin{figure*}[t]
  \centering
  \includegraphics[width=\textwidth]{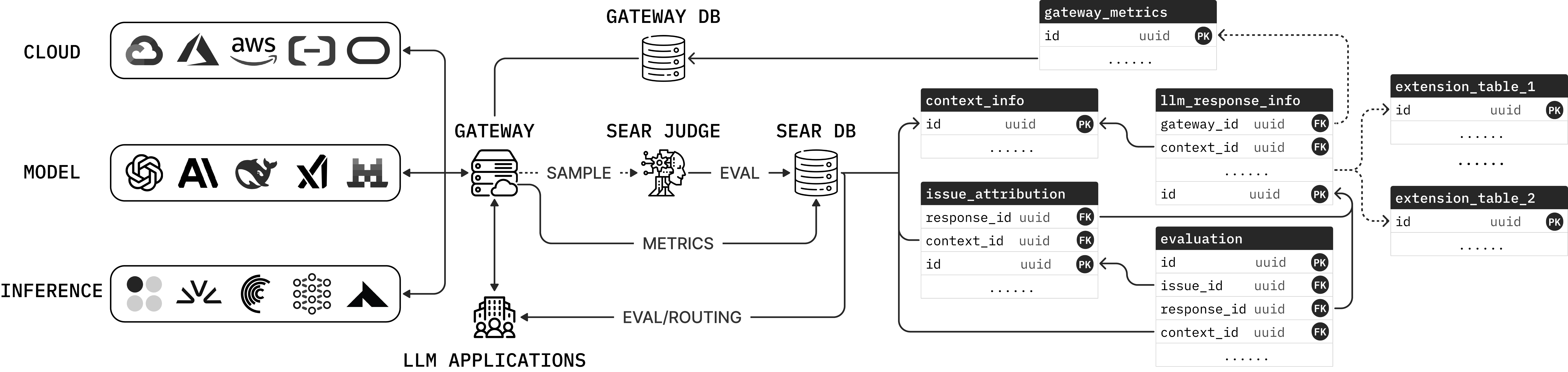}
  \caption{SEAR system architecture and database schema. A central gateway routes requests to LLM providers, samples traffic to the SEAR judge for evaluation, and logs operational metrics for all requests. Solid arrows denote mandatory foreign keys and dashed arrows denote optional foreign keys from the gateway metrics table.}
  \label{fig:schema-er}
\end{figure*}

SEAR extracts and reasons over LLM request sessions to produce
structured, typed evaluation records, which are co-located with
gateway operational metrics in a single SQL-queryable data layer.
We begin with a system overview, then describe the schema across
the semantic evaluation tables, the cross-table consistency design,
the gateway metrics table, and the extensibility mechanisms.

\subsection{Overview}
\label{sec:sear-overview}

Figure~\ref{fig:schema-er} illustrates the SEAR system architecture.
A central LLM gateway sits between LLM applications and providers. It
handles request routing, tracks prompt-cache usage, performs rate
limiting and failover, and logs operational metrics (latency, token
counts, cost, and error rates) for every request to the
\texttt{gateway\_metrics} table.
Because SEAR uses LLM-as-judge evaluation, scoring all traffic is
prohibitively expensive. The gateway therefore samples a configurable
fraction of requests for evaluation, where each sampled request
(referred to as a session) comprises the full conversation history
up to and including the current LLM response.

These sessions are then forwarded to the SEAR judge, a reasoning-LLM
judge that generates structured signals and inserts them into
a relational schema of multiple foreign-key-linked tables with
around one hundred typed columns spanning context signals, user
intent, response characteristics, issue attribution, and quality
scores.

By co-locating evaluation signals and gateway metrics in one queryable
data layer, downstream tasks such as routing, drift detection, and
provider benchmarking reduce to standard SQL queries over accumulated
records.
This forms a data flywheel where the gateway serves requests, sampled
sessions are logged and judged, and accumulated signals drive both
quality analysis for users and routing policy updates for subsequent
requests.

\subsection{Schema Design}
\label{sec:sear-database-schema}

The SQL-queryable data layer comprises five relational tables. The four
semantic evaluation tables are populated by the SEAR judge and
connected through cross-table consistency links. The gateway metrics
table is populated by the gateway for all request traffic.

\subsubsection{Semantic Evaluation Tables}
\label{sec:semantic-evaluation-tables}

The semantic evaluation layer consists of four relational tables, each
with typed columns (boolean, categorical enum, or ordinal) whose
per-table composition and type breakdown are summarized in
Table~\ref{tab:schema-summary}, with the full column-level schema and
foreign-key relationships in Appendix~\ref{app:full-schema}
(Figure~\ref{fig:full-schema}).
Each table targets a distinct stage of the LLM request lifecycle:

\begin{enumerate}
  \item \texttt{context\_info}:
    captures request-side context and intent from system messages,
    multi-turn user inputs, and reference material (e.g., language,
    domain, task type, and requirements such as tool use, code, or
    multi-step reasoning). In practice, context and intent are often
    mixed across turns, so the judge must disentangle what the model is
    asked to do from supporting context before assigning signals. The table
    also includes 20 static session features
    (modality flags, message counts, and per-role token counts) derived
    directly from raw logs.
  \item \texttt{llm\_response\_info}:
    captures what the model actually produced (e.g., tool invocation,
    code generation, reasoning behavior, refusal). Its overlap with
    table~(1) enables request--response gap analysis
    (e.g., code requested but not produced).
  \item \texttt{issue\_attribution}:
    for each shared dimension, attributes detected gaps to likely
    sources (user input, context, model behavior, or mixed causes),
    enabling targeted root-cause diagnosis.
  \item \texttt{evaluation}:
    assigns issue severity on a four-level ordinal scale and reports
    overall quality dimensions including relevance, completeness,
    coherence, instruction following, factual accuracy, safety, and
    overall response quality.
\end{enumerate}

All semantic evaluation columns use discrete types such as booleans,
categorical enums, or ordinal enums with explicit level definitions.
Integer and floating-point scores are avoided because prior work shows
that LLM judges can be unstable on wide numeric scales, with score
clustering, sensitivity to prompt wording, and weak discrimination
between nearby values (e.g., 90 vs.\ 93)~\cite{haldar2025ratingroulette}.
Discrete labels reduce this ambiguity by assigning each value a clear
semantic boundary (e.g., \texttt{true}/\texttt{false}), improving
consistency and reproducibility. Section~\ref{sec:self-contained-generation}
describes how each column definition is made self-contained with
explicit scope, assignment conditions, and edge cases to minimize
inter-column interference.

\subsubsection{Cross-Table Schema Design}
\label{sec:cross-table-design}

The mirrored and overlapping dimensions across these tables implement a
\emph{cross-table schema design}, in which semantically related columns across tables
are explicitly aligned.
In this design, \texttt{llm\_response\_info} mirrors relevant
requirement flags from \texttt{context\_info},
\texttt{issue\_attribution} assigns responsibility for detected gaps,
and \texttt{evaluation} scores gap severity.
For example, in the \texttt{tool\_call} signal family, the schema
records whether tool use was required, whether it was produced, who is
responsible for any issue (user, context, model, or mixed), and how
severe that issue is.

This design offers two benefits:
\begin{enumerate}[leftmargin=2.6em]
  \item \textbf{Consistency checks}: disagreements among semantically linked columns
    are detectable via table links, surfacing LLM judging errors and hallucinations.
    Flagged records can then be re-judged or removed.
  \item \textbf{Signal traceability}: each signal can be traced
    through all four tables, from whether it is requested, to whether
    it is produced, to who is responsible, to how severe the issue is.
\end{enumerate}

As all four tables are linked by foreign keys, consistency checks
reduce to standard SQL joins.
Code~\ref{lst:cross-table-check} in Appendix~\ref{app:cross-table-sql}
shows the query that detects violated records for the
\texttt{tool\_call} signal family.
Those records can then be re-judged with a stronger model or
filtered out.

\begin{table}[t]
\caption{Signal composition per semantic evaluation table.  Columns show the
  number of signals by type; \emph{Static} denotes columns extracted
  directly from raw session data (e.g., message counts, token counts
  per role) without the LLM judge.}
\label{tab:schema-summary}
\centering
\small
\begin{tabular}{lrrrr r r}
\toprule
\textbf{Table} & \textbf{Bool} & \textbf{Cat} & \textbf{Ord} & \textbf{Text} & \textbf{Static} & \textbf{Total} \\
\midrule
\texttt{context\_info}          & 17 & 4  & 2  & 2 & 20 & 45 \\
\texttt{llm\_response\_info}   & 15 & 1  & 2  & 1 &  0 & 19 \\
\texttt{issue\_attribution}    &  1 & 19 & 0  & 0 &  0 & 20 \\
\texttt{evaluation}            &  2 & 1  & 28 & 0 &  0 & 31 \\
\midrule
\textbf{Total}                 & 35 & 25 & 32 & 3 & 20 & 115 \\
\bottomrule
\end{tabular}
\end{table}

\subsubsection{Evaluation Schema Extensibility}
\label{sec:schema-extensibility}

The semantic evaluation schema supports two extension paths.

First, new tables can be added through optional foreign-key
links to existing ones, without modifying the core schema.
For example, a tool-call quality extension table can just link to
\texttt{llm\_response\_info} records where
\texttt{llm\_response\_has\_tool\_call} is true.
Records that do not need this extension simply leave the
foreign key as null.

Second, new independent signal columns can be appended to existing
tables.
Because added columns can affect generation stability and per-signal
accuracy, each extension should be designed and validated as described
in Section~\ref{sec:self-contained-generation}.

Together, these mechanisms let the schema evolve incrementally without
disrupting existing data or core LLM evaluation workflows.

\subsubsection{Gateway Metrics Table}
\label{sec:gateway-table}

The \texttt{gateway\_metrics} table is populated directly by the serving
infrastructure, not by the LLM judge, and records operational metrics
for every request handled by the LLM gateway.
Because this table is populated during online serving, it covers 100\%
of traffic.
Each record captures:
\begin{enumerate}
  \item \textbf{Request identity}: user, model, provider, and region
    identifiers, linked to the gateway metadata tables.
  \item \textbf{Performance metrics}: total round-trip latency, time to
    first token (TTFT), end-to-end throughput (total tokens divided by
    latency), and generation speed (completion tokens divided by
    decoding time after the first token).
  \item \textbf{Request status}: failure and timeout flags, with error
    type and message for failed requests.
  \item \textbf{Token usage}: prompt, completion, reasoning, and total
    token counts per request.
  \item \textbf{Cache usage}: cached prompt tokens, cache read input
    tokens, and cache creation input tokens, following token-level
    cache reporting adopted by major providers.
\end{enumerate}
The \texttt{gateway\_metrics} table is linked to
\texttt{llm\_response\_info} via a foreign key, enabling joins between
semantic evaluation signals and operational metrics.

On its own, this table supports aggregate operational observability
(e.g., global p95 latency, throughput, and failure rate) and
slice-level diagnostics by region, model, and provider. When joined
with evaluated sessions, it enables quality--operational analyses such
as quality--latency/cost/throughput/provider.

\section{Schema-Driven Judge}
\label{sec:schema-driven-judge}

\begin{figure}[t]
  \centering
  \includegraphics[width=0.7\linewidth]{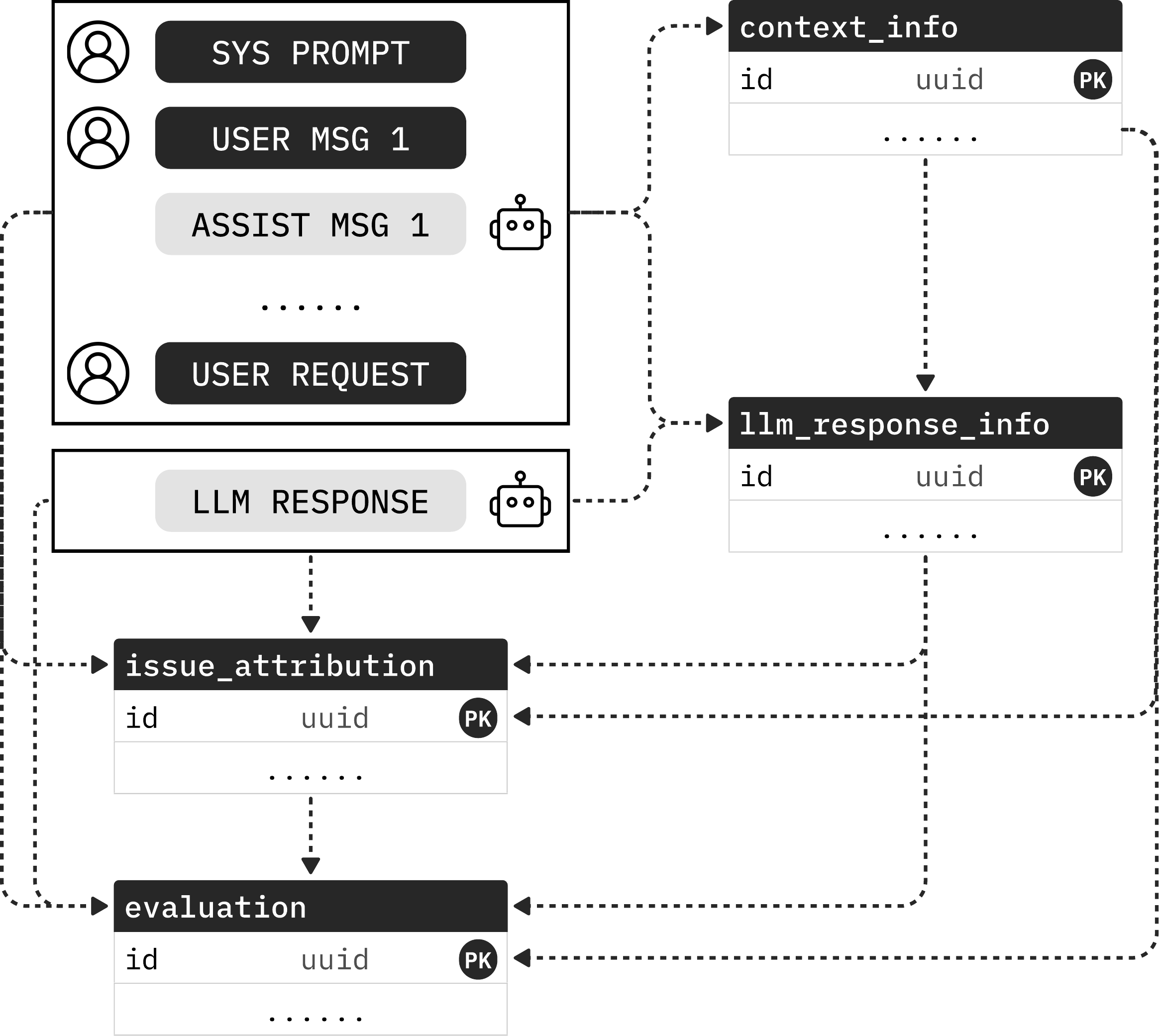}
  \caption{Each table's structured
    output call receives the conversation context and all upstream
    structured table outputs. Dashed arrows indicate input dependencies.}
  \label{fig:multi-stage-generation}
\end{figure}

Building on the schema design, SEAR must generate valid records for
those semantic evaluation tables with around one hundred
typed columns, far beyond the single-score or few-field outputs typical
of existing LLM judges.
The goal is to make schema-constrained generation at this scale and
complexity reliable in production settings.
This setting raises three challenges: (1) reliably producing large
structured outputs without inter-column confusion, (2) preserving
reasoning quality under strict schema constraints, and (3)
orchestrating generation across dependent tables efficiently.
This section addresses these challenges through self-contained signal
instructions, in-schema reasoning, and multi-stage generation.

\subsection{Self-Contained Signal Instructions}
\label{sec:self-contained-generation}

For each table, the judge emits all signals in a single structured
output call~\cite{openai2024structuredoutputs} by using a typed JSON
schema whose fields map directly to table columns.
This schema-to-output mapping avoids post-processing parsers and
produces records that are ready for direct insertion.
Code~\ref{lst:json-schema-sample} (Appendix~\ref{app:json-schema-sample})
shows an abbreviated schema.

An alternative is to have an LLM or coding agent generate SQL and then
execute ingestion logic. However, this adds an intermediate step with
extra token cost and a higher risk of type mismatches, syntax errors,
and security issues (Appendix~\ref{app:structured-vs-code}).

Unlike typical rubric-based evaluation, which often returns one or a
few fields per call~\cite{kim2024prometheus,zheng2023judging}, each
SEAR semantic evaluation table contains 15--30 typed fields with
potentially related semantics and complex cross-signal relationships.
Recent work shows that instruction-following compliance degrades
multiplicatively with the number of simultaneous
constraints~\cite{harada2025curseofinstructions}.
To reduce confusion between adjacent signals, we use a
\emph{self-contained instruction} design at the column level.
Each column description specifies its definition, evidence scope (which
input data to inspect), value-assignment rules, optional examples, and
edge cases that separate it from other columns.
This design reduces inter-column interference and helps the judge
evaluate and generate independent signals, as illustrated by the
\texttt{request\_requires\_tool\_call} field in
Code~\ref{lst:json-schema-sample} (Appendix~\ref{app:json-schema-sample}).

\subsection{In-Schema Reasoning}
\label{sec:in-schema-reasoning}

Self-contained signal instructions reduce inter-column confusion and
complex cross-signal relationships, but many fields still require
semantic reasoning (e.g., task type, domain category, and issue
attribution) rather than surface extraction.
Under strict output constraints, reasoning quality can degrade relative
to free-form generation~\cite{tam2024letmespeak}.
A natural approach is to prepend chain-of-thought
(CoT)~\cite{wei2022chain} reasoning in free text before
committing to structured fields.

Formally, let $x$ denote the input context (conversation history and
LLM response) and $Y = (Y_1, \dots, Y_d)$ the $d$ signal columns for
a given table.
A standard CoT approach generates a free-text reasoning trace $c$ in
a separate call, then conditions the structured output on it:
\begin{equation}
\label{eq:two-call-cot}
p(Y \mid x) = \sum_c p(Y \mid c, x)\, p(c \mid x)\,.
\end{equation}
However, this requires at least two LLM calls per table, one for $c$
and one for $Y$, doubling the total from four to eight calls for the
full schema.

We instead propose \emph{in-schema reasoning}: a temporary
\texttt{reasoning} text field $r$ placed as the first property in the
JSON schema (Code~\ref{lst:json-schema-sample}, Appendix~\ref{app:json-schema-sample}) and dropped before
database insertion.
Because generation follows schema order, the model emits $r$ before the
signal columns, yielding a single autoregressive pass:
\begin{equation}
\label{eq:in-schema-reasoning}
p(r, Y \mid x) = p(r \mid x) \prod_{i=1}^{d} p(Y_i \mid Y_{<i}, r, x)\,.
\end{equation}
The reasoning is therefore generated within the same structured
output call, requiring no additional LLM invocation.

We design the \texttt{reasoning} field description as a self-check
guidance prompt to improve structured signal-generation accuracy:
\begin{enumerate}
  \item \textbf{Identify the task.} State what the system and user
    messages ask the serving LLM to do, especially for complex requests.
  \item \textbf{Derive the following signals step by step.} Based on (1),
    determine which signal columns should be set and to what values.
  \item \textbf{Verify consistency.} Check that the derived signals are
    consistent with the stated task from (1).
\end{enumerate}
In practice, this guidance improves signal extraction and reasoning
during structured generation while preserving the single-call-per-table
design with minimal additional token cost.
Recent work on latent reasoning~\cite{hao2024coconut} suggests that
models benefit from attending to their own continuous hidden states
rather than re-encoded text, which may further explain the effectiveness
of single-pass in-schema reasoning.

\subsection{Multi-Stage Generation}
\label{sec:multi-stage-generation}

We orchestrate table generation in multiple stages, each with
carefully scoped context.
Figure~\ref{fig:multi-stage-generation} shows the stage dependencies
(dashed arrows). For example, the first stage
(\texttt{context\_info}) receives only the conversation context, and
each subsequent stage receives the same context plus all upstream
structured outputs.

The stage order follows request flow naturally: context and user intent
first, response characterization next, then issue attribution and
quality scoring. This sequence also aligns
with the foreign-key dependencies in the database schema
(Figure~\ref{fig:schema-er}).

This decomposition improves generation stability for two reasons. First,
each call emits a smaller schema (19--31 columns instead of around one
hundred in a single call), reducing malformed or inconsistent fields.
Second, recent evidence shows that longer context alone degrades LLM
performance even when all information is
relevant~\cite{du2025contextlength}; scoping each stage's input to its
immediate dependencies limits context length and reduces instruction
ambiguity.

Because each stage emits a schema-conforming JSON object for one table,
the process can materialize foreign-key links after all stages
complete and commit all linked records in one transaction, ensuring
cross-table consistency and simpler failure recovery.

\begin{table}[t]
  \centering
  \caption{Qualitative comparison of evaluation approaches.
    \cmark\ = natively supported, \pmark\ = partial,
    \xmark\ = not supported.}
  \label{tab:qualitative-comparison}
  \footnotesize
  \begin{tabular}{@{}lccccc@{}}
  \toprule
  \textbf{Capability} & \textbf{Ours} & \textbf{Text} & \textbf{Single} & \textbf{Rubric} & \textbf{Templ.} \\
  \midrule
  Root-cause attribution        & \cmark & \pmark & \xmark & \pmark & \pmark \\
  Quality scores                & \cmark & \pmark & \pmark & \cmark & \cmark \\
  Session aggregation           & \cmark & \pmark & \cmark & \cmark & \pmark \\
  Per-signal slice analysis     & \cmark & \pmark & \xmark & \pmark & \pmark \\
  Joint eval.\ + oper.\ metrics & \cmark & \pmark & \pmark & \pmark & \pmark \\
  Consistency checks            & \cmark & \xmark & \xmark & \xmark & \xmark \\
  \bottomrule
  \end{tabular}
\end{table}

\begin{table}[t]
  \centering
  \caption{Qualitative comparison of routing approaches.
    \cmark\ = natively supported, \pmark\ = partial,
    \xmark\ = not supported.}
  \label{tab:routing-comparison}
  \footnotesize
  \begin{tabular}{@{}lccc@{}}
  \toprule
  \textbf{Capability} & \textbf{Ours} & \textbf{Signal} & \textbf{Learned} \\
  \midrule
  Per-signal explanation         & \cmark & \pmark & \pmark \\
  Deep semantic extraction       & \cmark & \xmark & \pmark \\
  Preference optimization        & \cmark & \xmark & \cmark \\
  Quality--cost--latency tuning  & \cmark & \pmark & \pmark \\
  Oper.-metric-aware routing     & \cmark & \cmark & \pmark \\
  Joint evaluation and routing   & \cmark & \xmark & \xmark \\
  \bottomrule
  \end{tabular}
\end{table}

\section{Data-Driven Evaluation and Routing}
\label{sec:eval-routing}

Because SEAR records are structured and SQL-queryable, diverse
evaluation and routing workflows reduce to standard queries without
separate custom pipelines.
This section illustrates data-driven evaluation
(\S\ref{sec:data-driven-eval}) and routing (\S\ref{sec:routing}).
Tables~\ref{tab:qualitative-comparison}
and~\ref{tab:routing-comparison} compare SEAR with representative
alternatives.

\subsection{Data-Driven Evaluation}
\label{sec:data-driven-eval}

Because every signal is a typed, timestamped column, evaluation
reduces to SQL queries over the joint SEAR and gateway tables.
We illustrate with three scenarios.

\paragraph{Model Evaluation}
Code~\ref{lst:model-eval} (Appendix~\ref{app:eval-sql}) compares
candidate models on coding tasks over the most recent thirty days,
computing per-model LLM-caused issue rates and average severity by
domain.

\paragraph{Provider Evaluation}
Code~\ref{lst:provider-select} (Appendix~\ref{app:eval-sql})
ranks providers by task quality with median latency as a tiebreaker,
combining evaluation scores with gateway operational metrics.

\paragraph{User Evaluation}
Code~\ref{lst:user-eval} (Appendix~\ref{app:eval-sql}) profiles
per-user risk by aggregating safety-sensitive content, ambiguous
instructions, and noisy context across sessions. The gateway can use
these statistics to trigger guardrails or adjust sampling rates.

\subsection{Data-Driven Routing}
\label{sec:routing}

The same records support data-driven routing.
The router queries observed quality and cost across traffic slices to
produce model and provider routing strategies.

Existing routers extract request features via embeddings, classifiers,
or heuristics~\cite{ong2024routellm,ding2024hybrid,liu2025vllmsemanticrouter},
which is fast but limited to surface-level attributes.
SEAR instead derives signals through LLM reasoning over the full
request context, capturing complex semantics such as issue attribution
and cross-signal quality that shallow extractors miss, while producing
human-interpretable routing recommendations grounded in per-signal
evidence.
Here we illustrate three routing scenarios.

\paragraph{Model Routing}
Code~\ref{lst:model-route-cost} (Appendix~\ref{app:routing-sql})
selects the cheapest model whose quality is within 10\% of the best.
Code~\ref{lst:model-route-task} (Appendix~\ref{app:routing-sql})
targets a specific slice (creative tasks), keeping only models that
outperform the deployed one and ranking by cost.

\paragraph{Provider Routing}
The same model can behave differently across providers due to
differing serving setups.
Code~\ref{lst:provider-route} (Appendix~\ref{app:routing-sql})
selects providers whose quality is within 5\% of the best and ranks
them by median time to first token.

\paragraph{Real-Time Model Routing}
The above scenarios produce offline policies.
For real-time routing, a lightweight LLM classifies context-level
attributes (task type, complexity, domain) from the incoming request,
and the gateway looks up the matching SEAR-derived policy
(Algorithm~\ref{alg:realtime-routing}).

\begin{algorithm}[t]
\caption{Real-time routing via context classification.}
\label{alg:realtime-routing}
\KwIn{request $r$, routing policy table $P$, classifier model $M_c$, slice keys $K$}
\KwOut{target model $m$}
$\mathbf{s} \leftarrow M_c(r)$ \tcp*{classify context signals}
$\textit{slice} \leftarrow \{\mathbf{s}[k] : k \in K\}$ \tcp*{$K$ is configurable}
\eIf{$\textit{slice} \in P$}{
  $m \leftarrow P[\textit{slice}].\text{model}$\;
}{
  $m \leftarrow \text{default\_model}$\;
}
\Return{$m$}\;
\end{algorithm}

\section{Experiments}

\begin{table*}[t]
  \centering
  \caption{Overall judge metrics by model, reasoning effort (low/high),
    and in-schema reasoning (w/o = without, w/ = with). GPT-5.2 is
    reported only with in-schema reasoning.}
  \label{tab:overall-judge-metrics}
  \small
  \begin{tabular}{lccccccc}
  \hline
    & \multicolumn{4}{c}{GPT-5-mini} & & \multicolumn{2}{c}{GPT-5.2} \\
    \cline{2-5}\cline{7-8}
    Metric & Low (w/o) & Low (w/) & High (w/o) & High (w/) & & Low (w/) & High (w/) \\
    \hline
    Error rate $\downarrow$ & 0.1044 & 0.0989 & 0.0926 & 0.0916 & & 0.0965 & \textbf{0.0851} \\
    Hamming loss $\downarrow$ & 0.1003 & 0.0958 & 0.0878 & 0.0869 & & 0.0914 & \textbf{0.0794} \\
    Boolean accuracy $\uparrow$ & 0.9170 & 0.9184 & 0.9368 & 0.9376 & & 0.9354 & \textbf{0.9549} \\
    Boolean micro F1 $\uparrow$ & 0.8243 & 0.8292 & 0.8646 & 0.8680 & & 0.8510 & \textbf{0.8990} \\
    Categorical accuracy $\uparrow$ & 0.9223 & 0.9295 & 0.9237 & 0.9258 & & 0.9287 & \textbf{0.9337} \\
    Ordinal MAE $\downarrow$ & 0.2745 & 0.2663 & 0.2018 & \textbf{0.1957} & & 0.2551 & 0.2185 \\
    Ordinal RMSE $\downarrow$ & 0.5999 & 0.5814 & \textbf{0.5191} & 0.5293 & & 0.5767 & 0.5439 \\
    Ordinal norm.\ MAE $\downarrow$ & 0.0834 & 0.0805 & \textbf{0.0625} & 0.0627 & & 0.0778 & 0.0683 \\
    \hline
    \end{tabular}
\end{table*}

\begin{table}[t]
        \centering
        \caption{Model ranking on the simple-complexity slice by
          composite quality (sum of six ordinal signals, max\,=\,18).}
        \label{tab:routing-perf}
        \small
        \begin{tabular}{lccc}
        \hline
        Model & Quality & \$/M\textsubscript{in} & \$/M\textsubscript{out} \\
        \hline
        gemini-2.5-flash-lite     & 17.57 & 0.10 & 0.40 \\
        claude-haiku-4-5 (deployed) & 17.00 & 1.00 & 5.00 \\
        grok-4-1-fast             & 16.86 & 0.20 & 0.50 \\
        qwen3-80b                 & 15.66 & 0.15 & 1.20 \\
        \hline
        \end{tabular}
    \end{table}

\subsection{Setup}
\label{sec:setup}

\paragraph{Dataset.}
Unlike prior work that relies on public benchmarks susceptible to data
contamination~\cite{hasan2025pitfalls,banerjee2024vulnerability} and
limited production
representativeness~\cite{fodor2025benchmarklimitations}, we evaluate on
production traffic.
We randomly sample 3{,}000 sessions from three organizations
(1{,}000 each) with distinct workload profiles: multilingual,
roleplay, and translation-heavy, among others.
Sessions span single-turn requests and multi-turn agentic workflows.
We reserve 300 sessions (100 per organization) as a held-out test set,
annotated by two senior engineers across all semantic evaluation
columns.
We additionally use 60 validation sessions (20 per organization),
sampled outside the 3{,}000-session pool, for schema development.
Per-organization ground-truth distributions are provided in
Appendix~\ref{app:dataset}.
All sessions are sampled and analyzed only where permitted under each
organization's data-use, consent, and privacy policies.

\paragraph{Models.}
We use two OpenAI reasoning models as LLM judges: GPT-5-mini and
GPT-5.2.
Both expose a configurable reasoning-effort parameter; we evaluate low
and high effort for each.
All judge configurations share the same system prompt
(Appendix~\ref{app:judge-system-prompt}).
Unless otherwise noted, experiments outside the labeled test and
validation sets use GPT-5-mini with high reasoning effort and
in-schema reasoning.

\paragraph{Metrics.}
We report metrics grouped by signal type.
For boolean signals: accuracy and micro-averaged F1, which is robust
to class imbalance in production data.
For categorical signals: accuracy.
For ordinal signals (severity, quality, complexity, completeness,
hallucination risk): MAE, RMSE, and normalized MAE (MAE divided by
scale length).
We also report two aggregate metrics: error rate (fraction of
predictions differing from ground truth) and Hamming loss (mean
per-record error rate).

\paragraph{Routing.}
For the routing case study (Section~\ref{sec:routing-perf}), we define
composite quality as the sum of six ordinal evaluation signals, each on
a three-point scale (task-type quality, response completeness,
instruction-following, factual accuracy, response relevance, and
response coherence), yielding a score in $\{6,\dots,18\}$.
The case study targets Organization~C, where almost all tasks fall into
the simple-complexity slice and are served by claude-haiku-4-5
(\$1.00/M\textsubscript{in}, \$5.00/M\textsubscript{out}).

\begin{figure}[t]
  \centering
  \includegraphics[width=\linewidth]{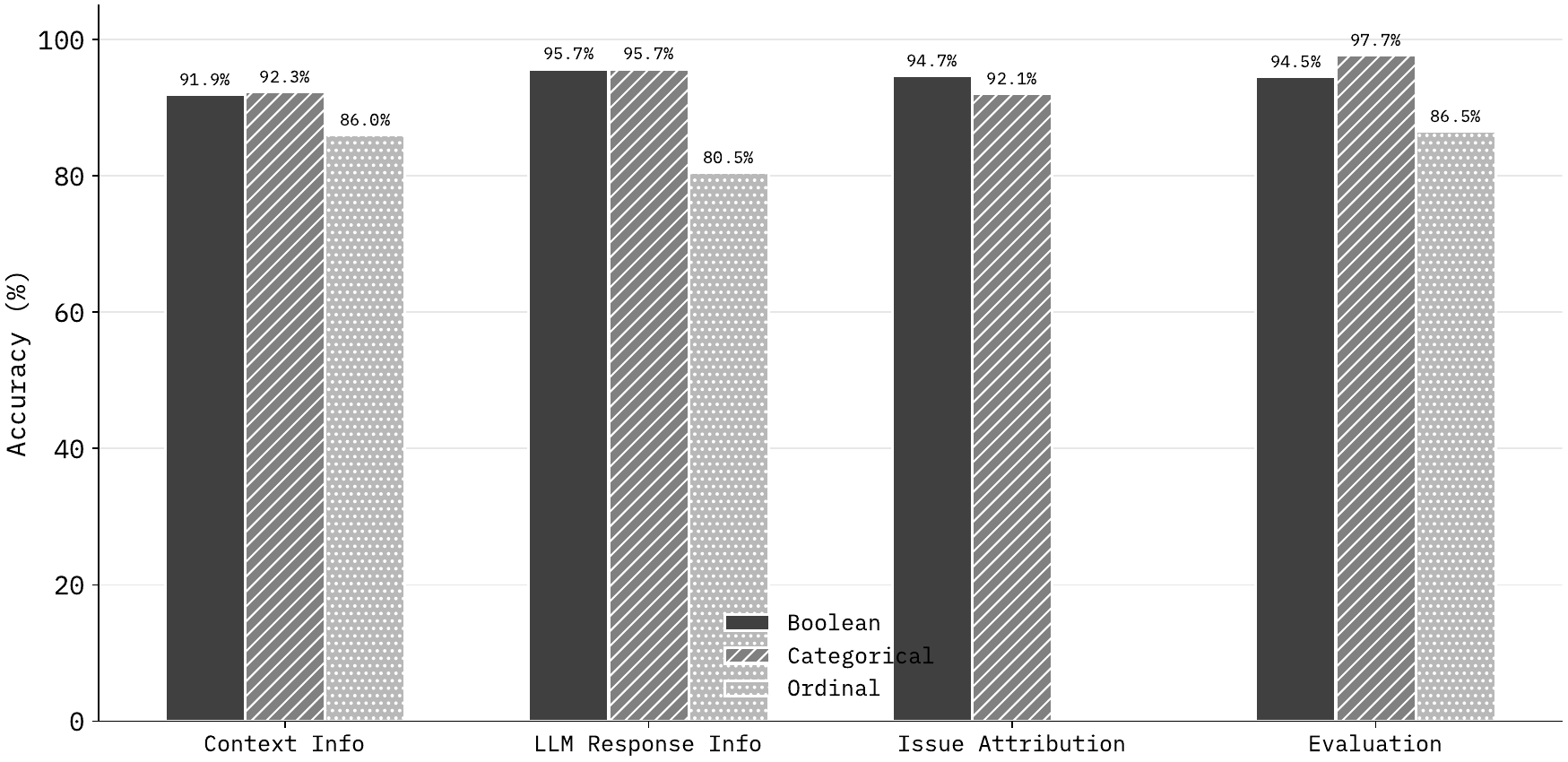}
  \caption{Per-table accuracy by signal type (GPT-5-mini, high effort, with in-schema reasoning).}
  \label{fig:accuracy-by-type}
\end{figure}

\subsection{Evaluation Performance}

Table~\ref{tab:overall-judge-metrics} reports aggregate judge metrics
across six configurations.
Overall, higher reasoning effort consistently improves accuracy, and
GPT-5.2 and GPT-5-mini at high effort achieve similar performance
across most metrics, with GPT-5.2 High leading on aggregate error rate,
boolean micro~F1, and categorical accuracy.
Figure~\ref{fig:accuracy-by-type} breaks down per-signal accuracy
under our default configuration (GPT-5-mini, high effort, in-schema
reasoning): boolean signals exceed 91\%, categorical signals exceed
92\%, and ordinal accuracy ranges from 80\% to 86\%.
We further ablate reasoning effort and in-schema reasoning in
Section~\ref{sec:ablation}.

\subsection{Routing Performance}
\label{sec:routing-perf}

Beyond judge accuracy, we evaluate whether SEAR signals can support
practical routing decisions: for a given workload slice, select the
lowest-cost model that maintains comparable quality.

\paragraph{Candidate selection.}
Using the composite quality metric defined in Section~\ref{sec:setup},
we query Organization~C sessions in the simple-complexity slice to rank
candidate models (query in Appendix~\ref{app:routing-sql},
Code~\ref{lst:routing-perf}).
The query filters out models with fewer than 10 sessions and ranks by
composite quality.
Table~\ref{tab:routing-perf} reports the results:
gemini-2.5-flash-lite achieves the highest quality score
(17.57) while costing 90\% less on input tokens and 92\% less on
output tokens than the deployed model.

\paragraph{Verification.}
To validate that SEAR-based routing preserves output quality, we replay
100 Organization~C sessions using gemini-2.5-flash-lite
(highest quality in Table~\ref{tab:routing-perf}) and
manually compare each response pair
(1 = routed model wins, 0.5 = tie, 0 = original wins).
Among the 100 sessions, 72 are ties, 12 favor
gemini-2.5-flash-lite, and 16 favor
claude-haiku-4-5, yielding a win rate of 48\%, effectively tied quality at 90\% lower input cost and 92\% lower
output cost.

\paragraph{Limitations.}
This case study covers one organization, one workload slice, and an
offline replay of 100 sessions, so it should not be interpreted as a
general routing benchmark.
Nevertheless, it demonstrates that SEAR can identify
substantially lower-cost candidates without measurable quality
degradation, even with limited logged traffic.

\subsection{Real-Time Context Classification}
\label{sec:realtime-classification}

The routing experiment in Section~\ref{sec:routing-perf} operates
offline. The SEAR judge evaluates sessions post-hoc, and routing
policies are derived from warehouse queries.
For real-time routing, the gateway needs to classify context-level
signals of incoming requests before they reach the target LLM so that
pre-computed routing policies can be applied at serving time
(Algorithm~\ref{alg:realtime-routing}).
This yields a two-tier architecture: (1)~a lightweight model classifies
context attributes inline at the gateway, and (2)~the gateway looks up
SEAR-derived routing policies keyed on those attributes to select the
target model, separating the heavy semantic evaluation (offline,
asynchronous) from the lightweight classification (online, per-request).
As a preliminary validation, we evaluate whether GPT-5-nano can
accurately extract context-level signals across the full
\texttt{context\_info} schema.

\paragraph{Setup.}
We run GPT-5-nano with minimal reasoning effort and in-schema reasoning
on the 300 human-labeled test sessions, prompting it to classify the
\texttt{context\_info} columns, the subset of SEAR signals that
describe the request rather than the response and are therefore
available pre-routing.
These include categorical, ordinal, and boolean signals.
We compare predictions against human annotations using the same metrics
as Table~\ref{tab:overall-judge-metrics}.

\paragraph{Results.}
Table~\ref{tab:realtime-context} reports classification accuracy on
the context-level signals.
GPT-5-nano achieves 82.6\% boolean accuracy, 72.3\% categorical
accuracy, and an ordinal MAE of 0.64.
We compare these results against the full SEAR judge and analyze
inference cost in Section~\ref{sec:ablation-realtime}.

\begin{table}[t]
  \centering
  \caption{Real-time context classification metrics for GPT-5-nano
    with minimal reasoning effort and in-schema reasoning, evaluated
    on the \texttt{context\_info} signals (23 columns).}
  \label{tab:realtime-context}
  \small
  \begin{tabular}{lc}
    \hline
    Metric & GPT-5-nano \\
    \hline
    Error rate $\downarrow$            & 0.2094 \\
    Hamming loss $\downarrow$          & 0.2094 \\
    Boolean accuracy $\uparrow$       & 0.8263 \\
    Boolean micro F1 $\uparrow$       & 0.6299 \\
    Categorical accuracy $\uparrow$   & 0.7233 \\
    Ordinal MAE $\downarrow$          & 0.6433 \\
    Ordinal RMSE $\downarrow$         & 0.9883 \\
    Ordinal norm.\ MAE $\downarrow$   & 0.1608 \\
    \hline
  \end{tabular}
\end{table}

\section{Ablation Studies}
\label{sec:ablation}

\subsection{Cross-Table Consistency Checks}
\label{sec:ablation-cross-table}

The cross-table consistency predicates
(Section~\ref{sec:cross-table-design}) enforce logical invariants
across the four evaluation tables via SQL-join rules.
For example, if both the request and response boolean flags for a
signal are false, the corresponding severity must be
\texttt{not\_applicable}.
Records that violate any predicate are filtered out, and all metrics
are recomputed on the consistent subset.

Table~\ref{tab:cross-table-ablation} reports the results.
Weaker configurations (lower reasoning effort or no in-schema
reasoning) produce more inconsistencies. GPT-5-mini low without
in-schema reasoning flags 34 records (11.3\%), whereas GPT-5-mini
high with in-schema reasoning flags only 2 (0.7\%).
Filtering yields the largest metric improvements for weaker
configurations (up to $+$7.5\% error rate reduction) but can
slightly over-prune stronger ones ($-$0.3\% for GPT-5-mini high
with in-schema reasoning), indicating diminishing returns when the
judge is already accurate.
Overall, cross-table consistency checking serves as a practical
post-hoc quality-assurance mechanism, most effective when judge
quality is lower.

\subsection{Multi-Stage vs.\ Single-Stage Judge}

We test whether the full SEAR schema (${\sim}100$ typed columns) can
be generated in a single structured-output call instead of the
four-stage generation (\S\ref{sec:multi-stage-generation}).
However, the single-stage configuration frequently produces malformed
or incomplete JSON, preventing reliable quantitative comparison.
In contrast, the multi-stage design generates one table per call,
each with a smaller output schema, and allows later stages to
reference upstream outputs for improved cross-table coherence.
These results confirm that multi-stage generation is necessary to
make schema-constrained evaluation practical at this scale.

\subsection{Reasoning Effort and In-Schema Reasoning}

As shown in Table~\ref{tab:overall-judge-metrics}, reasoning effort is
the dominant factor that increasing effort from low to high yields larger
gains than adding in-schema reasoning
(Section~\ref{sec:in-schema-reasoning}) at the same effort level.
This pattern is also reflected in GPT-5-mini at high effort with
in-schema reasoning outperforming GPT-5.2 at low effort on error rate,
indicating that higher reasoning effort can offset model-capability
gaps.
At the same time, in-schema reasoning remains complementary at both
low and high effort, enabling it consistently improves most metrics.
Together, strong schema-based evaluation performance depends on both
sufficient reasoning capability and effective in-schema reasoning.

\subsection{Meta-Task Confusion}

Across configurations, we identify a recurring failure mode that we
call \textit{meta-task confusion}: the judge conflates its own
schema-based evaluation instructions (e.g., JSON output requirements)
with the user's original task, producing false positives on
\texttt{context\_info} columns such as
\texttt{request\_requires\_output\_format}.
This is especially prevalent in production settings, where
instruction-dense context blurs the boundary between user intent and
judge framing.

We report this analysis only for configurations with in-schema
reasoning, as the generated reasoning chain is what makes meta-task
confusion identifiable in the output.
We manually verified all affected sessions.
Table~\ref{tab:meta-task-confusion-ablation} reports the confusion
rate by model and reasoning effort.
The rate decreases monotonically with stronger reasoning: from
7.3\% (GPT-5-mini low) to 0.0\% (GPT-5.2 high), confirming that
increased inference-time compute effectively mitigates this failure
mode.

\subsection{Real-Time Context Classification}
\label{sec:ablation-realtime}

Compared with the default SEAR judge (GPT-5-mini, high effort,
in-schema reasoning) on the same \texttt{context\_info} columns
(Section~\ref{sec:realtime-classification}), GPT-5-nano shows
higher error rate (0.21 vs.\ 0.09), lower boolean accuracy
(0.83 vs.\ 0.92, micro~F1 0.63 vs.\ 0.86), lower categorical
accuracy (0.72 vs.\ 0.91), and higher ordinal MAE
(0.64 vs.\ 0.22).
However, since only 1 of 4 tables is needed and GPT-5-nano with
minimal reasoning runs at ${\sim}\tfrac{1}{4}$ the latency of
GPT-5-mini, the end-to-end cost is ${\sim}\tfrac{1}{16}$ of the
full pipeline. In product we can restrict to the 3 routing-relevant columns
(task type, domain, complexity) to reduce the latency further.
Moreover, routing typically discretizes complexity into coarse
buckets (e.g., simple vs.\ non-simple), where adjacent-level
ordinal errors still map to the same routing decision, so the
higher ordinal MAE has limited impact on downstream routing
accuracy.

\begin{table}[t]
  \centering
  \caption{Cross-table consistency filtering.
    \emph{Incon.}\ is the number of inconsistent records.
    \emph{Imp.}\ columns show relative change after filtering
    (positive = improvement).}
  \label{tab:cross-table-ablation}
  \small
  \begin{tabular}{lrc rr}
  \toprule
  \textbf{Configuration} & \textbf{Incon.} & \textbf{Filter\%}
    & \textbf{Imp.\ Err $\uparrow$} & \textbf{Imp.\ HL $\uparrow$} \\
  \midrule
  GPT-5-mini low (w/o)  & 34 & 11.33\% & $+$7.47\% & $+$6.78\% \\
  GPT-5-mini low (w/)   &  9 &  3.00\% & $+$2.22\% & $+$2.09\% \\
  GPT-5-mini high (w/o) & 11 &  3.67\% & $+$0.86\% & $+$0.46\% \\
  GPT-5-mini high (w/)  &  2 &  0.67\% & $-$0.33\% & $-$0.35\% \\
  GPT-5.2 low (w/)      & 17 &  5.67\% & $+$4.56\% & $+$4.38\% \\
  GPT-5.2 high (w/)     &  6 &  2.00\% & $+$0.82\% & $+$0.76\% \\
  \bottomrule
  \end{tabular}
  \end{table}

  \begin{table}[t]
    \centering
    \caption{Meta-task confusion rate for schema-conforming judge runs
      (all with in-schema reasoning), by model and reasoning effort.}
    \label{tab:meta-task-confusion-ablation}
    \small
    \begin{tabular}{cccc}
    \hline
    \multicolumn{2}{c}{GPT-5-mini} & \multicolumn{2}{c}{GPT-5.2} \\
    \cline{1-2}\cline{3-4}
    Low & High & Low & High \\
    \hline
    7.3\% (22 / 300) & 3.7\% (11 / 300) & 2.0\% (6 / 300) & 0.0\% (0 / 300) \\
    \hline
    \end{tabular}
  \end{table}

\section{Future Work}

Our preliminary results show that the lightweight context classifier
can extract routing-relevant signals at low cost
(Section~\ref{sec:realtime-classification}). A next step is to conduct
end-to-end online routing experiments that combine this classifier with
organization-specific routing policies, measuring whether per-request
routing preserves downstream quality while providing clear,
schema-grounded explanations for each routing decision.

Data coverage is another limitation. Our current dataset covers a
limited set of models and task types, so routing experiments focus on
simple-complexity slices with sufficient cross-model overlap. We plan
to expand data collection to cover broader task and complexity ranges
and to evaluate routing over a wider model pool.

Finally, we plan to run the SEAR judge with models beyond the
GPT family, including both non-GPT closed-source models and
open-source models.

\section{Conclusion}

We presented SEAR, a schema-based evaluation and routing system for
multi-model, multi-provider LLM gateways.
SEAR defines an extensible relational schema with cross-table
consistency checks and uses self-contained signal instructions,
in-schema reasoning, and multi-stage generation to produce
database-ready structured outputs.
Because signals are derived through LLM reasoning rather than shallow
classifiers, SEAR captures complex request semantics, provides
human-interpretable routing explanations, and unifies evaluation and
routing in a single SQL-queryable layer alongside gateway operational
metrics.

Across 3{,}000 production sessions, the SEAR judge achieves high
accuracy on boolean, categorical, and ordinal signals against
human-labeled data.
In a routing case study, SEAR-derived queries identify a substitute
model with 90\% lower input cost and 92\% lower output cost while
maintaining comparable quality.
Beyond offline routing, a lightweight context classifier further
demonstrates that routing-relevant signals can be extracted at low
latency, enabling real-time, per-request routing.

\bibliographystyle{ACM-Reference-Format}
\bibliography{references}

\clearpage
\appendix
\appendix

\section{SEAR Full Database Schema}
\label{app:full-schema}

\begin{figure*}[htbp]
  \centering
  \includegraphics[width=0.9\textwidth]{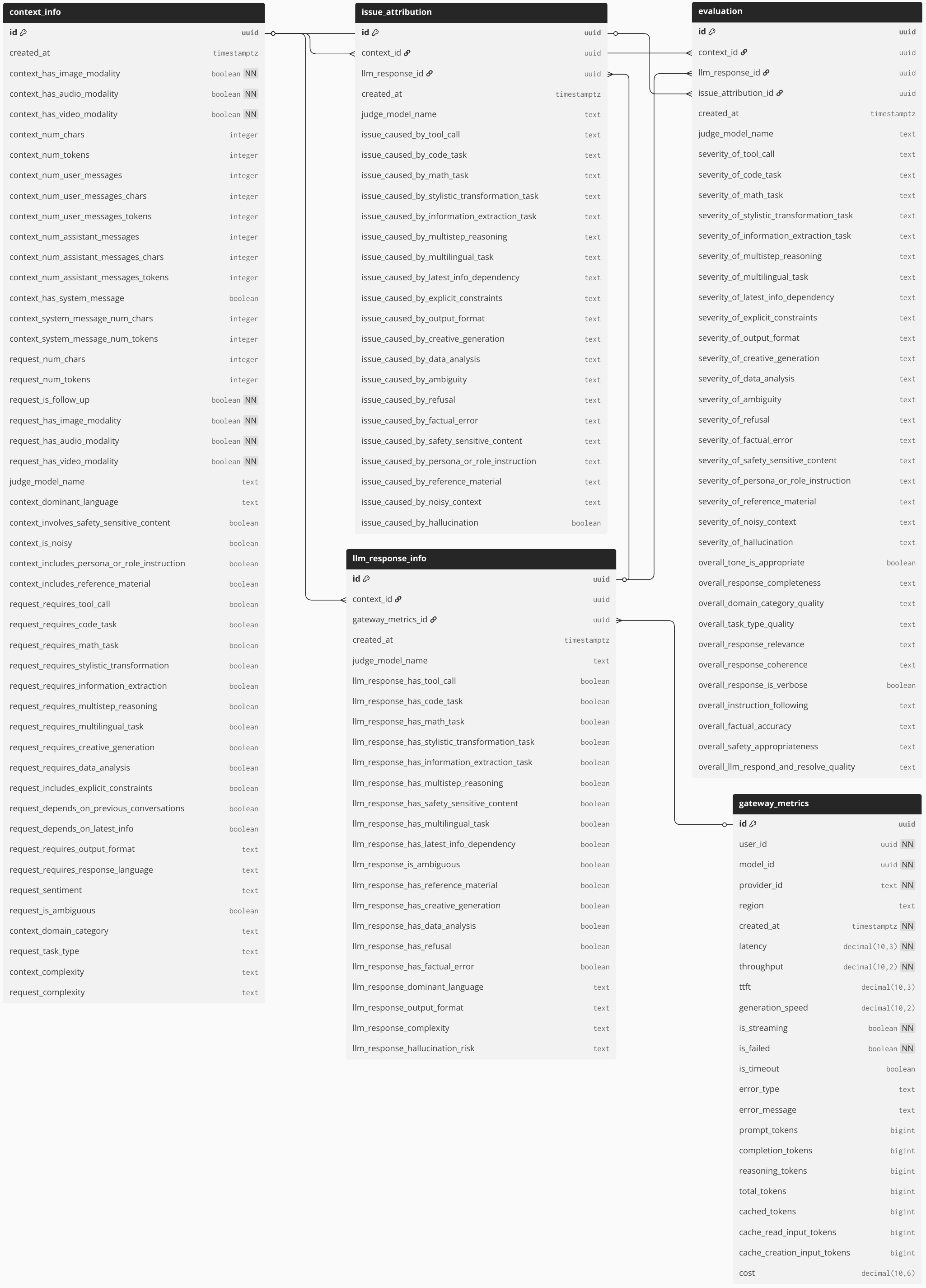}
  \caption{Full SEAR database schema including the four semantic evaluation tables and the gateway metrics table. Each table lists all columns with their types.}
  \label{fig:full-schema}
\end{figure*}

Figure~\ref{fig:full-schema} presents the full SEAR database schema,
including the four semantic evaluation tables and the gateway metrics table,
their columns and types, and the foreign-key relationships that link them.

\section{Cross-Table Consistency Query}
\label{app:cross-table-sql}

Code~\ref{lst:cross-table-check} detects cross-table consistency
violations for the \texttt{tool\_call} signal family, flagging absence
violations, issue--severity mismatches, and orphan severity entries.

\begin{listing}[htbp]
\caption{Cross-table violation detection for the \texttt{tool\_call}
  signal family.}
\label{lst:cross-table-check}
\begin{lstlisting}[language=SQL]
SELECT ctx.id, llm.id, ia.id, eval.id
FROM context_info      AS ctx
JOIN llm_response_info AS llm  ON llm.context_id = ctx.id
JOIN issue_attribution AS ia   ON ia.context_id  = ctx.id
JOIN evaluation        AS eval ON eval.context_id = ctx.id
WHERE
  (ctx.request_requires_tool_call    = FALSE
   AND llm.llm_response_has_tool_call = FALSE
   AND (ia.issue_caused_by_tool_call  NOT IN ('not_applicable','none')
     OR eval.severity_of_tool_call    NOT IN ('not_applicable','none')))
  OR (ia.issue_caused_by_tool_call   IN ('not_applicable','none')
   AND eval.severity_of_tool_call    NOT IN ('not_applicable','none'))
  OR (eval.severity_of_tool_call     IN ('minor','major')
   AND ia.issue_caused_by_tool_call  IN ('not_applicable','none'));
\end{lstlisting}
\end{listing}

\section{Example JSON Schema}
\label{app:json-schema-sample}

\begin{listing}[htbp]
\begin{lstlisting}
{
  "description": "Metadata, facts and insights from
    user messages for analytics and routing analysis.",
  "properties": {
    "reasoning": {
      "type": "string",
      "description": "Think step-by-step:
        (1) What is the system/user message instructing the LLM to do?
        (2) Based on (1), which fields should be set?
        (3) Verify conclusions against (1)."
    },
    "request_requires_tool_call": {
      "type": "boolean",
      "description": "Whether the user's intent requires
        the LLM to invoke external tools. ..."
    },
    "context_domain_category": {
      "type": "string",
      "enum": ["trivia", "marketing", "legal",
        "health", "academia", "finance", ...],
      "description": "Subject area of the conversation. ..."
    },
    ......
  }
}
\end{lstlisting}
\caption{Abbreviated \texttt{context\_info} JSON schema provided to the LLM judge for
   generation.}
\label{lst:json-schema-sample}
\end{listing}

\section{Structured Output vs.\ Code-Generation Approaches}
\label{app:structured-vs-code}

An alternative is to have the LLM generate SQL \texttt{INSERT}
statements instead of structured JSON.
This introduces three drawbacks: (1)~token overhead from SQL syntax,
(2)~code-generation errors orthogonal to evaluation (syntax errors,
type mismatches, malformed escaping), and (3)~security risks from
executing LLM-generated SQL in a production pipeline.
Structured output avoids all three: the LLM produces only typed
values, and an ORM inserts them into the database.

\section{Data-Driven Evaluation and Routing Examples}
\label{app:eval-examples}

\subsection{Evaluation Examples}
\label{app:eval-sql}

Code~\ref{lst:model-eval} compares candidate models on coding tasks in
the most recent 30 days by computing, for each model and domain, the
rate at which the LLM introduced issues and the average severity of
those issues.

\begin{listing}[htbp]
\caption{Model comparison on coding tasks over the most recent 30 days,
  by LLM-caused issue rate and severity grouped by domain.}
\label{lst:model-eval}
\begin{lstlisting}[language=SQL]
SELECT gw.model_id,
       ctx.context_domain_category  AS domain,
       COUNT(*)                     AS n,
       AVG(CASE WHEN ia.issue_caused_by_code_task
                  IN ('llm','both')
            THEN 1 ELSE 0 END)      AS llm_issue_rate,
       AVG(CASE eval.severity_of_code_task
             WHEN 'major' THEN 2 WHEN 'minor' THEN 1
             ELSE 0 END)            AS avg_severity
FROM context_info      ctx
JOIN issue_attribution ia   ON ia.context_id  = ctx.id
JOIN evaluation        eval ON eval.context_id = ctx.id
JOIN llm_response_info llm  ON llm.context_id  = ctx.id
JOIN gateway_metrics   gw
  ON gw.id = llm.gateway_metrics_id
WHERE ctx.request_task_type = 'coding'
  AND gw.created_at >= NOW() - INTERVAL '30 days'
GROUP BY 1, 2
ORDER BY llm_issue_rate, avg_severity;
\end{lstlisting}
\end{listing}

Code~\ref{lst:provider-select} ranks providers by joining evaluation
quality scores with gateway latency metrics.

\begin{listing}[htbp]
\caption{Provider ranking by task quality and median latency.}
\label{lst:provider-select}
\begin{lstlisting}[language=SQL]
SELECT gw.provider_id,
       PERCENTILE_CONT(0.5) WITHIN GROUP
         (ORDER BY gw.latency)       AS p50_latency,
       AVG(CASE eval.overall_task_type_quality
             WHEN 'high' THEN 3 WHEN 'medium' THEN 2
             WHEN 'low'  THEN 1 ELSE 0 END)
                                     AS avg_quality
FROM gateway_metrics   gw
JOIN llm_response_info llm
  ON llm.gateway_metrics_id = gw.id
JOIN evaluation        eval
  ON eval.context_id = llm.context_id
WHERE gw.is_failed = FALSE
GROUP BY 1
HAVING COUNT(*) >= 30
ORDER BY avg_quality DESC, p50_latency;
\end{lstlisting}
\end{listing}

Code~\ref{lst:user-eval} aggregates user-side risk
indicators---safety-sensitive content, ambiguous instructions, and noisy
context---across each user's sessions, surfacing users whose prompts
frequently pose evaluation challenges.

\begin{listing}[htbp]
\caption{Per-user risk profiling from context signals.}
\label{lst:user-eval}
\begin{lstlisting}[language=SQL]
SELECT gw.user_id,
       COUNT(*)                          AS n,
       AVG(ctx.context_involves_safety_sensitive_content
           ::int)                        AS safety_rate,
       AVG(ctx.request_is_ambiguous::int)
                                         AS ambiguity_rate,
       AVG(ctx.context_is_noisy::int)    AS noisy_rate
FROM context_info      ctx
JOIN llm_response_info llm
  ON llm.context_id = ctx.id
JOIN gateway_metrics   gw
  ON gw.id = llm.gateway_metrics_id
GROUP BY 1
HAVING COUNT(*) >= 10
ORDER BY safety_rate DESC, ambiguity_rate DESC;
\end{lstlisting}
\end{listing}

\subsection{Routing Examples}
\label{app:routing-sql}

Code~\ref{lst:model-route-cost} finds the cheapest model whose
aggregate quality is within 10\% of the best-performing model.
The \texttt{model\_provider} table is a gateway metadata table
that records per-million-token input and output costs for each
model--provider pair.

\begin{listing}[htbp]
\caption{Minimum-cost model within 10\% quality of the best.}
\label{lst:model-route-cost}
\begin{lstlisting}[language=SQL]
WITH model_perf AS (
  SELECT gw.model_id,
         AVG(CASE eval.overall_task_type_quality
               WHEN 'high' THEN 3 WHEN 'medium' THEN 2
               WHEN 'low'  THEN 1 ELSE 0 END)
                                        AS avg_quality,
         AVG(gw.prompt_tokens  * mp.input_cost_per_million_token
           + gw.completion_tokens
             * mp.output_cost_per_million_token) AS avg_cost
  FROM gateway_metrics   gw
  JOIN llm_response_info llm
    ON llm.gateway_metrics_id = gw.id
  JOIN evaluation        eval
    ON eval.context_id = llm.context_id
  JOIN model_provider mp
    ON mp.model_id = gw.model_id
   AND mp.provider_id = gw.provider_id
  WHERE gw.is_failed = FALSE
  GROUP BY 1
  HAVING COUNT(*) >= 30
)
SELECT model_id, avg_quality, avg_cost
FROM model_perf
WHERE avg_quality
   >= 0.9 * (SELECT MAX(avg_quality) FROM model_perf)
ORDER BY avg_cost;
\end{lstlisting}
\end{listing}

Code~\ref{lst:model-route-task} targets creative writing tasks,
finding models that strictly outperform the currently deployed model
and ranking them by cost.

\begin{listing}[htbp]
\caption{Models with strictly better quality on creative tasks,
  ranked by cost.}
\label{lst:model-route-task}
\begin{lstlisting}[language=SQL]
WITH task_perf AS (
  SELECT gw.model_id,
         AVG(CASE eval.overall_task_type_quality
               WHEN 'high' THEN 3 WHEN 'medium' THEN 2
               WHEN 'low'  THEN 1 ELSE 0 END)
                                        AS avg_quality,
         AVG(gw.prompt_tokens  * mp.input_cost_per_million_token
           + gw.completion_tokens
             * mp.output_cost_per_million_token) AS avg_cost
  FROM gateway_metrics   gw
  JOIN llm_response_info llm
    ON llm.gateway_metrics_id = gw.id
  JOIN evaluation        eval
    ON eval.context_id = llm.context_id
  JOIN context_info      ctx
    ON ctx.id = llm.context_id
  JOIN model_provider mp
    ON mp.model_id = gw.model_id
   AND mp.provider_id = gw.provider_id
  WHERE ctx.request_task_type = 'creative'
    AND gw.is_failed = FALSE
  GROUP BY 1
  HAVING COUNT(*) >= 30
)
SELECT model_id, avg_quality, avg_cost
FROM task_perf
WHERE avg_quality > (
    SELECT avg_quality FROM task_perf
    WHERE model_id = :current_model_id)
ORDER BY avg_cost;
\end{lstlisting}
\end{listing}

Code~\ref{lst:provider-route} selects, for a given model, the top
providers whose quality is within 5\% of the best provider, ranked by
median time to first token (TTFT).

\begin{listing}[htbp]
\caption{Top providers for a given model within 5\% quality,
  ranked by TTFT.}
\label{lst:provider-route}
\begin{lstlisting}[language=SQL]
WITH provider_perf AS (
  SELECT gw.provider_id,
         AVG(CASE eval.overall_task_type_quality
               WHEN 'high' THEN 3 WHEN 'medium' THEN 2
               WHEN 'low'  THEN 1 ELSE 0 END)
                                         AS avg_quality,
         PERCENTILE_CONT(0.5) WITHIN GROUP
           (ORDER BY gw.ttft)            AS p50_ttft
  FROM gateway_metrics   gw
  JOIN llm_response_info llm
    ON llm.gateway_metrics_id = gw.id
  JOIN evaluation        eval
    ON eval.context_id = llm.context_id
  WHERE gw.model_id = :target_model_id
    AND gw.is_failed = FALSE
  GROUP BY 1
  HAVING COUNT(*) >= 30
)
SELECT provider_id, avg_quality, p50_ttft
FROM provider_perf
WHERE avg_quality
   >= 0.95 * (SELECT MAX(avg_quality)
              FROM provider_perf)
ORDER BY p50_ttft
LIMIT 3;
\end{lstlisting}
\end{listing}

Code~\ref{lst:routing-perf} implements cost-aware model substitution:
it computes per-model quality on simple-complexity tasks and
ranks candidates by quality.

\begin{listing}[htbp]
\caption{Cost-aware model substitution: rank models by quality
  on simple-complexity tasks.}
\label{lst:routing-perf}
\begin{lstlisting}[language=SQL]
SELECT llm.model_name,
       COUNT(*)                          AS n,
       AVG(CASE eval.overall_task_type_quality
             WHEN 'high' THEN 3 WHEN 'medium' THEN 2
             WHEN 'low' THEN 1 ELSE 0 END
         + CASE eval.overall_response_completeness
             WHEN 'complete' THEN 3 WHEN 'partial' THEN 2
             WHEN 'incomplete' THEN 1 ELSE 0 END
         + CASE eval.overall_instruction_following
             WHEN 'high' THEN 3 WHEN 'medium' THEN 2
             WHEN 'low' THEN 1 ELSE 0 END
         + CASE eval.overall_factual_accuracy
             WHEN 'high' THEN 3 WHEN 'medium' THEN 2
             WHEN 'low' THEN 1 ELSE 3 END
         + CASE eval.overall_response_relevance
             WHEN 'high' THEN 3 WHEN 'medium' THEN 2
             WHEN 'low' THEN 1 ELSE 0 END
         + CASE eval.overall_response_coherence
             WHEN 'high' THEN 3 WHEN 'medium' THEN 2
             WHEN 'low' THEN 1 ELSE 0 END)
                                         AS avg_quality,
       MAX(mp.input_cost_per_million_token)
                                         AS input_price,
       MAX(mp.output_cost_per_million_token)
                                         AS output_price
FROM context_info        ctx
JOIN llm_response_info   llm  ON llm.context_id = ctx.id
JOIN evaluation          eval ON eval.context_id = ctx.id
JOIN gateway_metrics     gw   ON gw.id = llm.gateway_metrics_id
JOIN model_provider      mp   ON mp.model_id = gw.model_id
                              AND mp.provider_id = gw.provider_id
WHERE ctx.request_complexity = 'simple'
GROUP BY 1
HAVING COUNT(*) >= 10
ORDER BY avg_quality DESC;
\end{lstlisting}
\end{listing}

\section{Dataset}
\label{app:dataset}

Tables~\ref{tab:dataset-context-info}--\ref{tab:dataset-evaluation}
report the per-organization ground-truth distributions for all signals
in the 300-session test set.
Boolean signals show the True percentage; categorical and ordinal
signals show the dominant value and its percentage.

\begin{table*}[htbp]
  \centering
  \caption{Ground-truth distribution of \texttt{context\_info} signals (300 sessions, 100 per organization). Boolean signals show True\%; categorical/ordinal signals show the dominant value and its percentage.}
  \label{tab:dataset-context-info}
  \small
  \begin{tabular}{llcccc}
  \hline
  Signal & Type & Org\,A & Org\,B & Org\,C & All \\
  \hline
  safety\_sensitive\_content & bool & 1\% & 0\% & 0\% & 0\% \\
  noisy & bool & 16\% & 0\% & 0\% & 5\% \\
  persona\_or\_role\_instruction & bool & 100\% & 100\% & 100\% & 100\% \\
  reference\_material & bool & 90\% & 100\% & 100\% & 97\% \\
  tool\_call & bool & 1\% & 0\% & 0\% & 0\% \\
  code\_task & bool & 0\% & 0\% & 0\% & 0\% \\
  math\_task & bool & 0\% & 0\% & 0\% & 0\% \\
  stylistic\_transformation & bool & 29\% & 7\% & 100\% & 45\% \\
  information\_extraction & bool & 17\% & 5\% & 0\% & 7\% \\
  multistep\_reasoning & bool & 38\% & 38\% & 0\% & 25\% \\
  multilingual\_task & bool & 16\% & 0\% & 100\% & 39\% \\
  creative\_generation & bool & 24\% & 66\% & 0\% & 30\% \\
  data\_analysis & bool & 0\% & 0\% & 0\% & 0\% \\
  explicit\_constraints & bool & 100\% & 100\% & 100\% & 100\% \\
  previous\_conversations & bool & 6\% & 0\% & 0\% & 2\% \\
  latest\_info & bool & 2\% & 0\% & 0\% & 1\% \\
  ambiguous & bool & 10\% & 0\% & 0\% & 3\% \\
  \hline
  language & cat & en (63\%) & en (100\%) & zh (72\%) & en (63\%) \\
  output\_format & cat & json (40\%) & json (48\%) & plain text (100\%) & plain text (57\%) \\
  response\_language & cat & en (64\%) & en (100\%) & zh (95\%) & en (56\%) \\
  sentiment & ord & neutral (99\%) & neutral (100\%) & neutral (100\%) & neutral (100\%) \\
  category & cat & ent./roleplay (23\%) & ent./roleplay (100\%) & other (35\%) & ent./roleplay (41\%) \\
  task\_type & cat & transform. (24\%) & creative/plan. (33\%) & transform. (100\%) & transform. (42\%) \\
  context\_complexity & cat & moderate (46\%) & complex (81\%) & simple (91\%) & complex (40\%) \\
  request\_complexity & cat & complex (33\%) & complex (70\%) & simple (89\%) & simple (40\%) \\
  \hline
  \end{tabular}
\end{table*}

\begin{table*}[htbp]
  \centering
  \caption{Ground-truth distribution of \texttt{llm\_response\_info} signals. Format follows Table~\ref{tab:dataset-context-info}.}
  \label{tab:dataset-llm-response}
  \small
  \begin{tabular}{llcccc}
  \hline
  Signal & Type & Org\,A & Org\,B & Org\,C & All \\
  \hline
  tool\_call & bool & 0\% & 0\% & 0\% & 0\% \\
  code\_task & bool & 1\% & 0\% & 2\% & 1\% \\
  math\_task & bool & 0\% & 0\% & 0\% & 0\% \\
  stylistic\_transformation\_task & bool & 22\% & 7\% & 90\% & 40\% \\
  information\_extraction\_task & bool & 13\% & 0\% & 1\% & 5\% \\
  multistep\_reasoning & bool & 6\% & 14\% & 0\% & 7\% \\
  safety\_sensitive\_content & bool & 1\% & 0\% & 0\% & 0\% \\
  multilingual\_task & bool & 9\% & 2\% & 91\% & 34\% \\
  latest\_info\_dependency & bool & 4\% & 0\% & 0\% & 1\% \\
  ambiguous & bool & 0\% & 0\% & 0\% & 0\% \\
  reference\_material & bool & 60\% & 68\% & 100\% & 76\% \\
  creative\_generation & bool & 25\% & 52\% & 0\% & 26\% \\
  data\_analysis & bool & 0\% & 0\% & 0\% & 0\% \\
  refusal & bool & 2\% & 0\% & 0\% & 1\% \\
  factual\_error & bool & 1\% & 0\% & 0\% & 0\% \\
  \hline
  language & cat & en (69\%) & en (95\%) & zh (84\%) & en (58\%) \\
  format & cat & json (43\%) & json (84\%) & plain text (97\%) & plain text (44\%) \\
  complexity & cat & simple (44\%) & mod./complex (36\%) & trivial/simple (39\%) & trivial/simple (32\%) \\
  hallucination\_risk & ord & none (72\%) & none (63\%) & none (100\%) & none (78\%) \\
  \hline
  \end{tabular}
\end{table*}

\begin{table*}[htbp]
  \centering
  \caption{Ground-truth distribution of \texttt{issue\_attribution} signals. Format follows Table~\ref{tab:dataset-context-info}.}
  \label{tab:dataset-issue-attribution}
  \small
  \begin{tabular}{llcccc}
  \hline
  Signal & Type & Org\,A & Org\,B & Org\,C & All \\
  \hline
  hallucination & bool & 0\% & 0\% & 0\% & 0\% \\
  \hline
  tool\_call & ord & n/a (99\%) & n/a (100\%) & n/a (100\%) & n/a (100\%) \\
  code\_task & ord & n/a (100\%) & n/a (100\%) & n/a (100\%) & n/a (100\%) \\
  math\_task & ord & n/a (100\%) & n/a (100\%) & n/a (100\%) & n/a (100\%) \\
  stylistic\_transformation\_task & ord & n/a (74\%) & n/a (93\%) & none (100\%) & n/a (56\%) \\
  information\_extraction\_task & ord & n/a (82\%) & n/a (95\%) & n/a (100\%) & n/a (92\%) \\
  multistep\_reasoning & ord & n/a (62\%) & n/a (62\%) & n/a (100\%) & n/a (75\%) \\
  multilingual\_task & ord & n/a (84\%) & n/a (100\%) & none (100\%) & n/a (61\%) \\
  latest\_info\_dependency & ord & n/a (98\%) & n/a (100\%) & n/a (100\%) & n/a (99\%) \\
  explicit\_constraints & ord & none (63\%) & none (64\%) & none (100\%) & none (76\%) \\
  output\_format & ord & none (58\%) & n/a (52\%) & none (83\%) & none (62\%) \\
  creative\_generation & ord & n/a (77\%) & none (66\%) & n/a (100\%) & n/a (70\%) \\
  data\_analysis & ord & n/a (100\%) & n/a (100\%) & n/a (100\%) & n/a (100\%) \\
  ambiguity & ord & n/a (55\%) & n/a (100\%) & n/a (100\%) & n/a (85\%) \\
  refusal & ord & n/a (98\%) & n/a (100\%) & n/a (100\%) & n/a (99\%) \\
  factual\_error & ord & none (95\%) & none (100\%) & none (100\%) & none (98\%) \\
  safety\_sensitive\_content & ord & n/a (99\%) & n/a (100\%) & n/a (100\%) & n/a (100\%) \\
  persona\_or\_role\_instruction & ord & none (98\%) & none (100\%) & none (100\%) & none (99\%) \\
  reference\_material & ord & none (90\%) & none (100\%) & none (100\%) & none (97\%) \\
  noisy\_context & ord & n/a (92\%) & n/a (100\%) & n/a (100\%) & n/a (97\%) \\
  \hline
  \end{tabular}
\end{table*}

\begin{table*}[htbp]
  \centering
  \caption{Ground-truth distribution of \texttt{evaluation} signals. Format follows Table~\ref{tab:dataset-context-info}.}
  \label{tab:dataset-evaluation}
  \small
  \begin{tabular}{llcccc}
  \hline
  Signal & Type & Org\,A & Org\,B & Org\,C & All \\
  \hline
  appropriate & bool & 100\% & 100\% & 100\% & 100\% \\
  verbose & bool & 7\% & 17\% & 0\% & 8\% \\
  \hline
  tool\_call & ord & n/a (99\%) & n/a (100\%) & n/a (100\%) & n/a (100\%) \\
  code\_task & ord & n/a (100\%) & n/a (100\%) & n/a (100\%) & n/a (100\%) \\
  math\_task & ord & n/a (100\%) & n/a (100\%) & n/a (100\%) & n/a (100\%) \\
  stylistic\_transformation\_task & cat & n/a (71\%) & n/a (93\%) & none (100\%) & n/a (55\%) \\
  information\_extraction\_task & cat & n/a (81\%) & n/a (95\%) & n/a (100\%) & n/a (92\%) \\
  multistep\_reasoning & cat & n/a (57\%) & n/a (62\%) & n/a (100\%) & n/a (73\%) \\
  multilingual\_task & cat & n/a (82\%) & n/a (100\%) & none (100\%) & n/a (61\%) \\
  latest\_info\_dependency & ord & n/a (99\%) & n/a (100\%) & n/a (100\%) & n/a (100\%) \\
  explicit\_constraints & cat & none (62\%) & none (64\%) & none (100\%) & none (75\%) \\
  output\_format & cat & none (56\%) & n/a (52\%) & none (83\%) & none (61\%) \\
  creative\_generation & cat & n/a (76\%) & none (66\%) & n/a (100\%) & n/a (70\%) \\
  data\_analysis & ord & n/a (100\%) & n/a (100\%) & n/a (100\%) & n/a (100\%) \\
  ambiguity & cat & n/a (55\%) & n/a (100\%) & n/a (100\%) & n/a (85\%) \\
  refusal & ord & n/a (98\%) & n/a (100\%) & n/a (100\%) & n/a (99\%) \\
  factual\_error & cat & none (85\%) & none (100\%) & none (100\%) & none (95\%) \\
  safety\_sensitive\_content & ord & n/a (99\%) & n/a (100\%) & n/a (100\%) & n/a (100\%) \\
  persona\_or\_role\_instruction & ord & none (94\%) & none (100\%) & none (100\%) & none (98\%) \\
  reference\_material & cat & none (87\%) & none (100\%) & none (100\%) & none (96\%) \\
  noisy\_context & ord & n/a (90\%) & n/a (100\%) & n/a (100\%) & n/a (97\%) \\
  hallucination & ord & none (82\%) & none (100\%) & none (100\%) & none (94\%) \\
  completeness & cat & complete (66\%) & complete (69\%) & complete (100\%) & complete (78\%) \\
  domain\_category\_quality & cat & high (80\%) & high (100\%) & high (100\%) & high (93\%) \\
  task\_type\_quality & cat & high (53\%) & high (58\%) & high (100\%) & high (70\%) \\
  relevance & cat & high (85\%) & high (98\%) & high (100\%) & high (94\%) \\
  coherence & cat & high (94\%) & high (87\%) & high (100\%) & high (94\%) \\
  instruction\_following & cat & high (54\%) & high (63\%) & high (100\%) & high (72\%) \\
  factual\_accuracy & cat & none (73\%) & none (100\%) & none (58\%) & none (77\%) \\
  safety\_appropriateness & ord & appropriate (99\%) & appropriate (100\%) & appropriate (100\%) & appropriate (100\%) \\
  respond\_and\_resolve\_quality & cat & high (52\%) & high (58\%) & high (100\%) & high (70\%) \\
  \hline
  \end{tabular}
\end{table*}

\section{Judge System Prompt}
\label{app:judge-system-prompt}

Code~\ref{lst:judge-system-prompt} shows the system prompt used by
the SEAR judge.

\begin{listing}[htbp]
\caption{SEAR system prompt used for schema-conforming table generation.}
\label{lst:judge-system-prompt}
\begin{lstlisting}
You are an evaluation assistant. Given the raw
session input and upstream structured outputs (if provided), produce one
JSON object that exactly matches the target table schema.

Follow these rules:
1. Focus on the target table for each stage and use upstream structured
   outputs as supporting evidence.
2. Follow each field description as the self-contained instruction
   including definition, scope, value conditions, and edge cases.
3. Set values only based on evidence in the raw input, tool results, or
   upstream structured outputs. If evidence is insufficient, use the
   conservative value defined by the schema (e.g., none or
   not_applicable).
4. If a reasoning field is present, use it to identify the task, derive
   fields step by step, and self-check consistency.
5. Raw input is provided as <input>...</input>.
\end{lstlisting}
\end{listing}

\end{document}